# Response to Nauenberg's "Critique of *Quantum Enigma: Physics Encounters Consciousness*"

Fred Kuttner
Department of Physics
University of California, Santa Cruz, CA 95064

## Abstract

Nauenberg's extended critique of *Quantum Enigma* rests on fundamental misunderstandings.

In his brief abstract, as a summary of his extensive paper[1], Michael Nauenberg *incorrectly* states that the "central claim" of our book, *Quantum Enigma*[2], is that "understanding quantum mechanics requires a conscious observer." In fact, we are explicit that understanding quantum mechanics, for all practical purposes, need *not* address the issue of consciousness. We rather note that physics has *encountered* consciousness. The major theme of Nauenberg's critique is that we are wrong, even doing something improper, by raising the issue of consciousness in connection with quantum mechanics. We must reply to this charge.

A dictionary's first definition of "encounter" is: "to meet, usually unexpectedly." It fits our use of the word. One such meeting early on was von Neumann's demonstration that, while for all practical purposes a wavefunction can be *considered* collapsed at any macroscopic point in the measurement chain, nevertheless, in principle no physical system described by quantum mechanics can collapse a wavefunction. The final collapse must take place at the level of consciousness[3]. We might also cite Wigner's famous comment that "…it was not possible to formulate the laws of quantum mechanics in a fully consistent way without reference to the consciousness."[4]

More recently a foremost exponent of "decoherence" in the measurement process, Zurek, has written: "An exhaustive answer to this question [the perception of a unique reality, i.e., a measurement] would undoubtedly have to involve a model of 'consciousness'…"[5] And in their discussion of the quantum potential interpretation of quantum mechanics, Bohm and Hiley write: "However, the intuition that consciousness and quantum theory are in some sense related seems to be a good one…"[6] Many examples where quantum mechanics has led physicists to speculate about a connection with consciousness could be cited.

And, of course, the connection has influenced philosophers. For example, Chalmers' landmark book[7] introducing the now much-discussed "hard problem" of consciousness has a final chapter titled "The Interpretation of Quantum Mechanics." With quantum mechanics, physics has at the very least *encountered* consciousness.

The tack we take in our book is to present the undisputed experimental facts with a quantum-theory-neutral demonstration. We have presented a technical version of such a quantum-theory-neutral demonstration some years ago.[8] Our book presents this to the general reader with the invitation to readers to decide on the extent of the encounter with consciousness for themselves. We present the quantum theory explanation of these demonstrations. But we leave the issue an unresolved mystery, an enigma, one that should stimulate meaningful and disciplined speculation.

In our book we are explicit that the encounter of physics with consciousness likely has no practical consequences for physics. It is metaphysics. Nauenberg criticizes us for talking of "metaphysics," as if metaphysics were pseudoscience. A major point of our book is that quantum mechanics brings us to an encounter with something beyond what physicists usually think of as physics. Something beyond physics is essentially the definition of metaphysics. We are clear in our book, that physicists, *as physicists*, need not concern themselves with consciousness. But we, and our readers, are more than just physicists. There is more to life than physics. Quantum mechanics tells of something mysterious that seems beyond "physics."

Exploring beyond testable physics, several interpretations of the meaning of quantum mechanics currently contend with the Copenhagen interpretation--which we all use in our teaching and research. In our book we treat nine of these (and in the paperback version planned by Oxford University Press, we treat ten). Many of these interpretations explicitly treat consciousness. For example, the most famous alternative to the Copenhagen interpretation, the "many worlds" interpretation, has also been seen as the "many minds" interpretation. Even when consciousness is not explicitly addressed, every interpretation has speculative implications for the nature of consciousness.

We are acutely aware that the strange implications of consciousness have increasingly been exploited to promote quantum nonsense. We not only consider this a serious societal problem, but we feel it to be the responsibility of physicists to address it. In fact, evading the enigma, or worse, denying it, cedes the field to the field to the purveyors of pseudoscience. We have urged our colleagues to teach the quantum mysteries honestly as an antidote to their misuse.[9]

Nauenberg refers to our "misunderstanding" of the physics and does so with very extensive quotations. We will not reply in detail to his many incorrect claims that we have the physics wrong. We just note that our book has been extensively reviewed and praised by physicists, and, other than Nauenberg's misinterpretations, no errors were noted. See our book's website, www.quantumenigma.com, for many reviews.

Some physicists can be unsettled by having our "hard" discipline of physics connected with the mysterious, "soft," and emotional subject of consciousness. Historian of science Jed Buchwald has noted that: "Physicists…have long had a special loathing for admitting questions with the slightest emotional content into their professional work."[10] At times we can, to an extent, share this reaction of our fellow physicists, but we are trying, along with many experts in the fundamentals of quantum theory, to move beyond it.

Early on John Bell wrote that it is likely that "the new way of seeing things will involve an imaginative leap that will astonish us."[11] Bell expressed similar views in what was probably the last article he ever published.[12] By "us" Bell did not just mean physicists. The astonishment Bell refers to might well involve consciousness.